\documentstyle[aps]{revtex}
\begin{document}
\author{S. V. Kuplevakhsky, A. V. Naduev,$^1$ and S. V. Naydenov$^2$}
\address{$^1$Department of Physics, Kharkov State University,\\
310077 Kharkov, Ukraine\\
$^2$Institute for Single Crystals, 310164 Kharkov, Ukraine}
\title{"Current-carrying states in superconductor/insulator and
superconductor/semiconductor superlattices in the mesoscopic regime". }
\maketitle

\begin{abstract}
We discuss some of the basic theoretical aspects of current-carrying states
in superconducting superlattices with tunnel barriers in the mesoscopic
regime, when $p_0^{-1}\ll a\ll \xi _0$ ($a$ is the superconducting layer
thickness, $p_0$ is the Fermi momentum, $\xi _0$ is the BCS coherence
length, $\hbar =1$). We establish the necessary conditions for the
observation of the classical Josephson effect (with sinusoidal current-phase
dependence) and derive self-consistent analytical expressions for the
critical Josephson current. These expressions are proportional to the small
factor $a/\xi _0$ and have unusual temperature dependence as compared with
the single-junction case. For certain parameter values, the superconducting
gap exhibits an exponential decrease due to pair-breaking effect of the
supercurrent. The supercurrent can completely destroy the superconductivity
of the system above a certain characteristic temperature $T^{*}$. In this
paper, we also study the effect of intrabarrier exchange interactions. We
show that this effect is strongly enhanced compared with the single-junction
case and can manifest itself in an exponential decrease of the critical
temperature.

\bigskip\ 

PACS numbers: 74.50.+r, 70.80.Dm
\end{abstract}

\newpage

\section{Introduction}

During the last years there have been major improvements in fabrication of
high-quality vertically stacked superconducting periodic multilayers
(superlattices) with Josephson coupling through insulating\cite{N93} and
semiconducting\cite{V97} barriers. These devices are promising candidates
for a variety of applications in microelectronics as, for example, high
frequency oscillators, mixers or fast switches.\cite{L94,LRK94,TUK97} On the
other hand, artificial Josephson coupled superlattices may serve\cite
{KMKPS94,TKU98} to model the properties of naturally layered high-$T_c$
superconductors exhibiting the intrinsic Josephson effect.\cite{KSKM92}

It is well-known \cite{BP82} that the principal physical characteristic of
any weakly-coupled superconducting system is the critical Josephson current, 
$j_c$. Unfortunately, the problem of self-consistent microscopic calculation
of $j_c$ in weakly-coupled superconducting\ superlattices poses serious
mathematical difficulties and was first addressed only recently.\cite
{KN97,KNG97} However, the very first theoretical results\cite{KN97,KNG97}
revealed dramatic differences with respect to single-junction behavior.
Whereas in the limit of thick superconducting (S) layers $a\gg \xi _0$ ($a$
for the S-layer thickness, $\xi _0$ for the BCS coherence length) the
description of the Josephson current in superlattices with tunnel barriers
converges with that in single junctions, in the opposite {\it mesoscopic}
regime $p_0^{-1}\ll a\ll \xi _0$ ($p_0$ is the Fermi momentum, $\hbar =1$)
drastic qualitative deviations from a single-junction case were found: a
strong reduction of $j_c$ due to the factor $a/\xi _0$, unusual temperature
dependence, and the suppression of the gap parameter in the S-layers by the
supercurrent. It should be emphasized that these subtle physical effects,
arising due to nonlocality of $j_c$, are not seized by simple
phenomenological models of layered superconductors, like, for example, the
wide-spread Lawrence-Doniach model,\cite{LD} that start with the local
description from the very beginning.

In this paper, we discuss the newly predicted nonlocal effects in mesoscopic
superlattices with tunnel barriers in more detail. In Section II, we present
a brief mathematical formulation of the problem. In Section III, we consider
superlattices with finite-size nonmagnetic insulating and semiconducting
barriers. In Section IV, we investigate the influence of intrabarrier
exchange interactions. Finally, in Section V, we present a summary of the
results and make some concluding remarks.

\section{Mathematical formalism}

Let us first formulate precisely the problem we are going to consider.
Throughout the paper, we study infinite periodic in the $x$-direction
superconducting ($s$-wave) systems with interlayer Josephson coupling in the
clean limit and in the absence of external magnetic fields. Complete
structural homogeneity in the $yz$-plane is implied, though the transverse
dimensions of the systems are taken to be small compared to the London
penetration depth in order to discard the influence of self-induced fields
and reduce the problem to effectively one-dimensional. The S-layers and the
barriers occupy the regions $S_{n}=\left[ -a-d/2+nc,-d/2+nc\right] $ and $%
B_{n}=\left[ -d/2+nc,d/2+nc\right] $, respectively ($c=a+d$ is the period,
and $n$ is an integer). Under these conditions, the problem is described
mathematically by the Gor'kov equations (Fourier-transformed in $y,z$):

\[
\left\{ i\omega +\left[ E_Ft^2+\frac 1{2m}\frac{d^2}{dx^2}-\hat U%
_B(x)\right] \tau _3+\frac i2(\tau _1+i\tau _2)\sigma _2\Delta (x)\right. 
\]
\begin{equation}  \label{1}
\left. -\frac i2(\tau _1-i\tau _2)\sigma _2\Delta ^{*}(x)\right\} \left[ 
\begin{array}{c}
\hat G_\omega (x,x^{\prime };t) \\ 
\hat F_\omega (x,x^{\prime };t)
\end{array}
\right] =\left[ 
\begin{array}{c}
\delta (x-x^{\prime }) \\ 
0
\end{array}
\right] ,
\end{equation}

\begin{equation}
\Delta ^{*}(x)=-\frac{i}{2}\left| g(x)\right| \pi N(0)v_{0}T\sum_{\omega
}\int_{0}^{1}dtt%
%TCIMACRO{\limfunc{tr}}
%BeginExpansion
\mathop{\rm tr}%
%EndExpansion
\left\langle \hat{F}_{\omega }(x,x;t)\sigma _{2}\right\rangle ,  \label{2}
\end{equation}

\[
g(x)=-|g|\sum_{n}\delta _{S_{n}}(x),
\]

\[
\delta _{\Omega }(x)=\left\{ 1,\text{ for }x\in \Omega ;0,\text{ for }%
x\notin \Omega \right\} .
\]
Here, $\hbar =1$, $\tau _{i}$ ($i=1,2,3$) are the Pauli matrices in the
Gor'kov-Nambu space, $\sigma _{i}$ ($i=1,2,3$) are the Pauli matrices in the
spin space, $\omega =\pi T(2n+1)$ ($n$ is an integer), $E_{F}$ is the Fermi
energy, $N(0)=mp_{0}/2\pi ^{2}$ is the one-spin density of states at the
Fermi level ($p_{0}=mv_{0}$ being the Fermi momentum), $g$ is the
electron-electron coupling constant, $t\equiv \cos \theta $ is the cosine of
the angle of incidence at the interface. The barrier potential is given by $%
\hat{U}_{B}(x)$, where the accent (\symbol{94}) denotes a non-trivial matrix
structure in the spin space. [In the absence of intrabarrier exchange
interactions $\hat{U}_{B\alpha \beta }(x)=U_{B}(x)\delta _{\alpha \beta }$.]
The functions $\hat{G}_{\omega }$, $\hat{F}_{\omega }$ are 2x2 matrices in
the spin space. In the self-consistency equation (\ref{2}), the trace is
taken over the spin indices, and $\left\langle \ldots \right\rangle $
denotes spatial averaging over atomic-scale oscillations. (We confine
ourselves to the limit $p_{0}^{-1}<<a$). Because of the periodicity, the
pair-potential obeys the relation $\Delta (x+nc)=\Delta (x)\exp (in\phi )$.
The functions $\hat{G}_{\omega }$, $\hat{F}_{\omega }$ and their first
derivatives $\hat{G}_{\omega }^{\prime }$, $\hat{F}_{\omega }^{\prime }$ are
subject to the usual continuity conditions at the interfaces $x=\pm d/2+nc.$

Our main interest is focused on the supercurrent density. For this quantity,
we employ the integral representation\cite{KNG97}

\[
j\equiv j(x\in B_0) 
\]

\[
=-2\pi ev_{0}N(0)T\sum_{\omega }\int_{0}^{1}dtt\int_{-\infty
}^{-d/2}dx_{1}\int_{d/2}^{\infty }dx_{2}\left[ \left\langle 
%TCIMACRO{\limfunc{tr}}
%BeginExpansion
\mathop{\rm tr}%
%EndExpansion
\left[ \hat{G}_{\omega }^{n}(x_{1},x_{2};t)\sigma _{2}\hat{G}_{-\omega
}^{t}(x_{2},x_{1};t)\sigma _{2}\right] \right\rangle \right. 
\]

\begin{equation}
\left. \times 
%TCIMACRO{\limfunc{Im}}
%BeginExpansion
\mathop{\rm Im}%
%EndExpansion
\left[ \Delta (x_{1})\Delta ^{*}(x_{2})\right] \right] ,  \label{3}
\end{equation}
where $\hat{G}_{\omega }^{n}$ is the Green's function of the system in the
normal state, and the upper index ($^{t}$) means transposition in the spin
space. An obvious advantage of this representation is the proportionality of
the integrand to the product of the Green's functions with $x_{1}\in S_{m}$, 
$x_{2}\in S_{n}$, where $m\neq n$, identically equal to zero in the absence
of weak coupling: To evaluate $j_{c}$ in first order in the tunneling
probability, $D$, we must take $\Delta $ in zero order and substitute the
expressions for $\hat{G}_{\omega }^{n}$, $\hat{G}_{-\omega }^{t}$ in first
order in $\sqrt{D}$. In this manner, we can circumvent a very difficult
problem of finding the full Green's function and calculate only physically
relevant elements of the latter, entering Eq. (\ref{3}), on the basis of a
perturbation expansion for Eqs. (\ref{1}), (\ref{2}) and the boundary
conditions.

\section{Superconductor/insulator (S/I) and superconductor/semiconductor
(S/Sem) superlattices with nonmagnetic barriers}

We begin by considering the case of a periodic structure with a finite-size
nonmagnetic repulsive barrier of the form 
\begin{equation}
\hat{U}_{B\alpha \beta }(x)=U_{0}\delta _{\alpha \beta }\sum_{n}\delta
_{B_{n}}(x),\text{ }U_{0}>0,  \label{4}
\end{equation}
which is typical of superconductor/semiconductor (S/Sem) multilayers. For
the potential (\ref{4}), the dependence of functions $\hat{G}_{\omega }$, $%
\hat{F}_{\omega }$ on the spin indices is trivial, 
\[
\left[ \hat{G}_{\omega }\right] _{\alpha \beta }=G_{\omega }\delta _{a\beta
},\text{ }\left[ \hat{F}_{\omega }\right] _{\alpha \beta }=F_{\omega
}i\sigma _{2\alpha \beta },
\]
and can be suppressed in what follows.

As we are not concerned with the Green's functions with coordinates inside
the barriers, we shall consider only $G_\omega (x\in S_n,x^{\prime }\in
S_m;t)$, $F_\omega (x\in S_n,x^{\prime }\in S_m;t)$. To derive the boundary
conditions for these functions, we first solve (\ref{1}) for $G_\omega (x\in
B_n,x^{\prime }\in S_m;t)$ and $F_\omega (x\in B_n,x^{\prime }\in S_m;t)$.
Making use of the full set of the boundary conditions, we arrive at the
required relations:

\[
\left[ 
\begin{array}{c}
G_\omega \\ 
F_\omega
\end{array}
\right] (nc+d/2,x^{\prime })=\cosh \left( \lambda _B^{\mp }d\right) \left[ 
\begin{array}{c}
G_\omega \\ 
F_\omega
\end{array}
\right] (nc-d/2,x^{\prime }) 
\]

\begin{equation}  \label{5}
+\frac{\sinh \left( \lambda _B^{\mp }d\right) }{\lambda _B^{\mp }}\left[ 
\begin{array}{c}
G_\omega ^{\prime } \\ 
F_\omega ^{\prime }
\end{array}
\right] (nc-d/2,x^{\prime }),
\end{equation}

\[
\left[ 
\begin{array}{c}
G_\omega ^{\prime } \\ 
F_\omega ^{\prime }
\end{array}
\right] (nc+d/2,x^{\prime })=\cosh \left( \lambda _B^{\mp }d\right) \left[ 
\begin{array}{c}
G_\omega ^{\prime } \\ 
F_\omega ^{\prime }
\end{array}
\right] (nc-d/2,x^{\prime }) 
\]

\begin{equation}  \label{6}
+\lambda _B^{\mp }\sinh \left( \lambda _B^{\mp }d\right) \left[ 
\begin{array}{c}
G_\omega \\ 
F_\omega
\end{array}
\right] (nc-d/2,x^{\prime }),
\end{equation}
where $\lambda _B^{\mp }=\sqrt{2m\left( U_0-E_Ft^2\mp i\omega \right) }$.
Equations (\ref{5}), (\ref{6}) form a closed system of exact boundary
conditions for the functions $G_\omega (x\in S_n,x^{\prime }\in S_m;t)$, $%
F_\omega (x\in S_n,x^{\prime }\in S_m;t).$ In the limit $d\rightarrow +0,$ $%
U_0\rightarrow +\infty $, $dU_0\equiv V=%
%TCIMACRO{\limfunc{const}}
%BeginExpansion
\mathop{\rm const}%
%EndExpansion
$, they reduce to the boundary conditions for a periodic delta-function
potential.

Assuming $\lambda _{B}^{\mp }\approx \lambda _{B}\equiv \sqrt{2m\left(
U_{0}-E_{F}t^{2}\right) }$, we proceed to the limit of a low-transparency
barrier $\lambda _{B}d>>1$. We can now solve (\ref{1}), (\ref{2}) with the
boundary conditions (\ref{5}), (\ref{6}) by means of perturbation theory,
with $\exp (-\lambda _{B}d)\propto \sqrt{D}$ being the expansion parameter.
As expected, in zero order, $\Delta ^{(0)}(x)=\Delta _{0}(T)\sum_{n}\exp
(in\phi )\delta _{S_{n}}(x)$ ($\Delta _{0}$ is the gap in the bulk, $\phi $
is a phase shift at the interfaces), and only $G_{\omega }^{(0)}(x\in
S_{n},x^{\prime }\in S_{n};t)$ and $F_{\omega }^{(0)}(x\in S_{n},x^{\prime
}\in S_{n};t)$ are nonzero. The functions $G_{\omega }(x\in S_{n},x^{\prime
}\in S_{m};t)$, $F_{\omega }(x\in S_{n},x^{\prime }\in S_{m};t)$ with $%
|n-m|>1$ are of order $>1$ in $\sqrt{D}$ and should be neglected. The
first-order approximation to $G_{\omega }(x\in S_{1},x^{\prime }\in S_{0};t)$%
, entering (\ref{1}), is given by

\[
G_\omega ^{(1)}(x\in S_1,x^{\prime }\in S_0;t)= 
\]

\[
=-\frac{m\lambda _Be^{-\lambda _Bd}}{\Omega ^2}\left\{ \frac{(\Omega +\omega
)^2+\Delta _0^2e^{i\phi }}{\lambda _{+}^2+\lambda _B^2}\frac{\sin \left[
\lambda _{+}(a+d/2-x+\alpha _{+})\right] \sin \left[ \lambda
_{+}(a+d/2+x^{\prime }+\alpha _{+})\right] }{\sin {}^2\left[ \lambda
_{+}(a+\beta _{+})\right] }\right. 
\]

\[
+\frac{\Delta _0^2\left( 1-e^{i\phi }\right) }{\lambda _{+}^2+\lambda _B^2} 
\sqrt{\frac{\lambda _{+}^2+\lambda _B^2}{\lambda _{-}^2+\lambda _B^2}}\frac{%
\sin \left[ \lambda _{+}(a+d/2-x+\alpha _{+})\right] \sin \left[ \lambda
_{-}(a+d/2+x^{\prime }+\alpha _{-})\right] }{\sin \left[ \lambda
_{+}(a+\beta _{+})\right] \sin \left[ \lambda _{-}(a+\beta _{-})\right] } 
\]

\[
+\frac{\Delta _0^2\left( 1-e^{i\phi }\right) }{\lambda _{-}^2+\lambda _B^2} 
\sqrt{\frac{\lambda _{-}^2+\lambda _B^2}{\lambda _{+}^2+\lambda _B^2}}\frac{%
\sin \left[ \lambda _{-}(a+d/2-x+\alpha _{-})\right] \sin \left[ \lambda
_{+}(a+d/2+x^{\prime }+\alpha _{+})\right] }{\sin \left[ \lambda
_{+}(a+\beta _{+})\right] \sin \left[ \lambda _{-}(a+\beta _{-})\right] } 
\]

\begin{equation}  \label{7}
+\left. \frac{(\Omega -\omega )^2+\Delta _0^2e^{i\phi }}{\lambda
_{-}^2+\lambda _B^2}\frac{\sin \left[ \lambda _{-}(a+d/2-x+\alpha
_{-})\right] \sin \left[ \lambda _{-}(a+d/2+x^{\prime }+\alpha _{-})\right] 
}{\sin {}^2\left[ \lambda _{-}(a+\beta _{-})\right] }\right\} ,
\end{equation}
where $\Omega =\sqrt{\omega ^2+\Delta _0^2}$, $\lambda _{\pm }=\pm \sqrt{%
2m\left( E_Ft^2\pm i\Omega \right) }$, $\alpha _{\pm }=\lambda _{\pm
}^{-1}\arcsin \left[ \lambda _{\pm }\left( \lambda _{\pm }^2+\lambda
_B^2\right) ^{-1/2}\right] $, and $\beta _{\pm }=\lambda _{\pm }^{-1}\arcsin
\left[ 2\lambda _{\pm }\lambda _B\left( \lambda _{\pm }^2+\lambda
_B^2\right) ^{-1/2}\right] $. The function $G_\omega ^{(1)}(x\in
S_0,x^{\prime }\in S_1;t)$, also entering (\ref{1}), is obtained from (\ref
{7}) via the substitution $x\leftrightarrow x^{\prime }$, $\phi \rightarrow
-\phi $. The normal-state function $G_\omega ^{n(1)}(x\in S_1,x^{\prime }\in
S_0;t)$ is a limiting case of (\ref{7}) for $\Delta _0=0$. Note that the
spatial dependence of (\ref{7}) clearly indicates that in first order in $D$
only two adjacent S-layers contribute to the supercurrent (\ref{3}).

Now one can benefit from the quasiclassical approximation $\lambda _{\pm
}\approx \pm p_{0}|t|+i\Omega /v_{0}|t|$ [$|t|>>(T_{c0}/E_{F})^{1/2}]$.
Inserting $\Delta ^{(0)}$ and quasiclassical expressions for $G_{\omega
}^{n(1)}$, $G_{\omega }^{(1)}$ into (\ref{3}), carrying out small-scale
averaging and performing spatial integration finally yields 
\begin{equation}
j=\frac{\pi }{2}ev_{0}N(0)\Delta _{0}^{2}T\int_{0}^{1}dttD(t)\sum_{\omega }%
\frac{\tanh \frac{2a\sqrt{\omega ^{2}+\Delta _{0}^{2}}}{v_{0}t}}{\omega
^{2}+\Delta _{0}^{2}}\sin \phi ,  \label{8}
\end{equation}

\[
D(t)=\frac{16E_Ft^2\left( U_0-E_Ft^2\right) }{U_0^2}\exp \left[ -2d\sqrt{%
2m\left( U_0-E_Ft^2\right) }\right] . 
\]
Equation (\ref{8}) is the desired analytical expression for the dc Josephson
current in clean S/I- and S/Sem-multilayers, valid for any $p_0^{-1}<<a\leq
\infty $ and arbitrary temperatures. [Concerning certain limitations of the
application of Eq. (\ref{8}) to {\it mesoscopic} superlattices, see the
discussion of Eq. (\ref{10}) below.] We observe that (\ref{8}) does not
depend on concrete features of our model and should also hold for
semiconducting barriers with internal structures of the type considered by
Aslamazov and Fistul in the case of a single SSemS-junction\cite{AF81}.
Equation (\ref{8}) is independent of the period $c$. This is a direct
consequence of the already-mentioned adjacent-layer coupling in first order
in $D$. The period may enter higher-order corrections, when the effect of
subgap current-carrying Bloch states\cite{KF92} comes into play.

As expected, in the limit $a>>\xi _0$, (\ref{8}) goes over into the
Ambegaokar-Baratoff relation\cite{AB63}

\begin{equation}
j=\frac{\pi ev_{0}N(0)}{2}\int_{0}^{1}dttD(t)\Delta _{0}(T)\tanh \frac{%
\Delta _{0}(T)}{2T}\sin \phi .  \label{9}
\end{equation}

However, the most interesting is the {\it mesoscopic regime}, when $%
p_0^{-1}<<a<<\xi _0.$ In this regime, using the gap equation for a bulk
superconductor, we arrive at the fundamental result

\begin{equation}
j=\frac{1}{2\pi }\frac{ev_{0}N(0)\Delta _{0}^{2}(T)}{T_{c0}}\frac{a}{\xi _{0}%
}\frac{\int_{0}^{1}dtD(t)}{\left| g\right| N(0)}\sin \phi ,  \label{10}
\end{equation}
where $\xi _{0}\equiv v_{0}/2\pi T_{c0}$.

First we note that the form of expression (\ref{10}) implies that in the
mesoscopic regime the true expansion parameter for the Josephson current is
the ratio $D/\left| g\right| N(0)\ll 1$ instead of $D\ll 1$ in the
Ambegaokar-Baratoff regime. [This conjecture is verified below. See the
discussion of relation (\ref{15}).] Taking account of the fact that in most
low-$T_c$ superconductors $\left| g\right| N(0)$ is typically $<$$0.3$, we
infer that the classical Josephson effects in mesoscopic superlattices can
be observed only under severe restrictions on the upper bound for $D$.

Three other remarkable features are to be noted with regard to (\ref{10}).
(i) A strong reduction of $j_{c}$ due to the emergence of the additional
small factor, $a/\xi _{0}$. (ii) The temperature dependence is determined
solely by the factor $\Delta _{0}^{2}(T)$ in the whole temperature range.
(iii) The occurrence of $\int_{0}^{1}dtD(t)$ instead of $\int_{0}^{1}dttD(t)$
in the Ambegaokar-Baratoff regime. These results is a manifestation of the
nonlocality of the supercurrent in its extreme: While the product of two
Green's functions in the integrand of (\ref{3}) decays at distances on the
order of $\xi _{0}$, the actual range of spatial integration is restricted
by two adjacent S-layers only. Thus, for instance, at $T=0$ we can rewrite (%
\ref{10}) as

\[
j=ev_{0}N(0)\Delta _{0}(0)\frac{\int_{0}^{1}dttD(t)P(t)}{\left| g\right| N(0)%
}\sin \phi ,
\]
where $P(t)=a/v_{0}t\Delta _{0}^{-1}(0)$ is the quasiclassical probability%
\cite{LL} of finding an unscattered electron with the $x$-component of the
velocity $v_{0}t$ within one S-layer during the characteristic time $\Delta
_{0}^{-1}(0)$. Finally, one must bear in mind that (\ref{10}) has been
derived in the clean limit. Considerable changes may occur in the dirty
limit $l<<\xi _{0}$ ($l$ is the electron mean free path), when the fall-off
length of the integrand in (\ref{3}) is of the order of $\sqrt{\xi _{0}l}$.
Actually, this limiting situation asks for further investigation.

Now we shall study the effect of suppression of the gap parameter by the
supercurrent in the mesoscopic regime, and establish the exact domain of
validity of Eq. (\ref{10}), obtained by perturbation methods. To this end,
we should consider the linearized self-consistency equation (\ref{2}). For
the sake of simplicity, we restrict ourselves to the case of thin ($d<<\min
\{a,\xi _0\}$) insulating repulsive barriers. For such barriers, the sought
equation, expanded to first order in $D<<1$, reads:

\[
\Delta (x\in S_{0})=\frac{\pi N(0)\left| g\right| }{v_{0}}\left[
\int_{-a+0}^{-0}dx^{\prime }\Delta (x^{\prime })T\sum_{\omega }\int_{0}^{1}%
\frac{dt}{t}\left\{ \exp \left[ -\frac{2|\omega |}{v_{0}t}\left| x-x^{\prime
}\right| \right] \right. \right. 
\]

\[
\left. \left[ 1-D(t)\right] \frac{\exp \left[ -\frac{2|\omega |a}{v_0t}%
\right] \cosh \left[ \frac{2|\omega |}{v_0t}\left( x-x^{\prime }\right)
\right] +\cosh \left[ \frac{2|\omega |}{v_0t}\left( x+x^{\prime }+a\right)
\right] }{\sinh \frac{2|\omega |a}{v_0t}}\right\} 
\]

\[
+\int_{+0}^{a-0}dx^{\prime }\Delta (x^{\prime })T\sum_{\omega }\int_{0}^{1}%
\frac{dt}{t}D(t)\left\{ \frac{\exp \left[ -\frac{2|\omega |}{v_{0}t}\left(
x^{\prime }-x-a\right) \right] }{2\cosh \frac{2|\omega |a}{v_{0}t}}\right. 
\]

\[
\left. +\frac{\exp \left[ -\frac{2|\omega |a}{v_0t}\right] \cosh \left[ 
\frac{2|\omega |}{v_0t}\left( x-x^{\prime }+a\right) \right] +\cosh \left[ 
\frac{2|\omega |}{v_0t}\left( x+x^{\prime }\right) \right] }{\sinh \frac{%
4|\omega |a}{v_0t}}\right\} 
\]

\[
+\int_{-2a+0}^{-a-0}dx^{\prime }\Delta (x^{\prime })T\sum_{\omega
}\int_{0}^{1}\frac{dt}{t}D(t)\left\{ \frac{\exp \left[ -\frac{2|\omega |}{%
v_{0}t}\left( x-x^{\prime }-a\right) \right] }{2\cosh \frac{2|\omega |a}{%
v_{0}t}}\right. 
\]

\begin{equation}
\left. \left. +\frac{\exp \left[ -\frac{2|\omega |a}{v_{0}t}\right] \cosh
\left[ \frac{2|\omega |}{v_{0}t}\left( x-x^{\prime }-a\right) \right] +\cosh
\left[ \frac{2|\omega |}{v_{0}t}\left( x+x^{\prime }-2a\right) \right] }{%
\sinh \frac{4|\omega |a}{v_{0}t}}\right\} \right] ,  \label{11}
\end{equation}
with a cutoff at the Debye frequency, $\omega _{D}$, in the sum over $\omega 
$ implied. The restriction on spatial integration by two adjacent S-layers
is again a result of the first-order approximation. For $a>>\xi _{0}$, (\ref
{11}) goes over into the very familiar equation for a single SIS-junction,%
\cite{dG89} describing spatial dependence of $\Delta $ near $T_{c0}$ in the
vicinity of the barrier. The analysis of (\ref{11}) in the limit $%
p_{0}^{-1}<<a<<\xi _{0}$ shows that in the current-carrying state it has
non-trivial solutions for $T<T_{c0}$. These solutions are given by a complex 
$\Delta $, constant in each S-layer with a phase shift $\phi _{c}$ at the
interfaces. (Physically, but for the phase jumps $\phi $ no spatial
variations can occur in the small-scale-averaged $\Delta $ over distances
less than $\xi _{0}$.) Thus, substituting $\Delta (x)=\Delta
(T)\sum_{n=-1}^{1}\exp (in\phi _{c})\delta _{S_{n}}(x)$ into Eq. (\ref{11})
yields 
\begin{equation}
\left| \phi _{c}\right| =\arccos \left[ 1+\frac{|g|N(0)}{\int_{0}^{1}dtD(t)}%
\ln \frac{T}{T_{c0}}\right] ,  \label{12}
\end{equation}
where $T^{*}\leq T\leq T_{c0}$, with 
\begin{equation}
T^{*}=T_{c0}\exp \left[ -\frac{1}{|g|N(0)}\int_{0}^{1}dtD(t)\right] .
\label{13}
\end{equation}
The physical meaning of relations (\ref{12}), (\ref{13}) is unraveled by the
fact that in the temperature range $T^{*}\leq T\leq T_{c0}$ the
superconductivity of the S-layers completely vanish due to depairing effect
of the supercurrent, when the phase difference reaches the critical value $%
\left| \phi \right| =\left| \phi _{c}\right| $. In this sense, the critical
phase difference $\left| \phi _{c}\right| $ in mesoscopic superlattices
plays the same role as the critical superfluid velocity\cite{dG89,T96} $%
v_{c}\sim \Delta (T)/p_{0}$ in thin superconducting wires.

At temperatures $T<T^{*}$, the Josephson current cannot completely destroy
the superconductivity of S-layers, but still one has to face strong
suppression of the gap parameter, unless the condition $D/\left| g\right|
N(0)\ll 1$ is fulfilled. Thus, at $T=0$ we have 
\begin{equation}
|\Delta (0)|=\Delta _{0}(0)\exp \left[ -\frac{1}{|g|N(0)}\int_{0}^{1}dtD(t)%
\left( 1-\cos \phi \right) \right] ,  \label{14}
\end{equation}
where $\Delta _{0}(0)=2\omega _{D}\exp \left[ -\frac{1}{|g|N(0)}\right] $ is
the bulk gap at $T=0$. In this case, the influence of the Josephson current
can even be regarded as effective weakening of the electron-electron
coupling constant: 
\[
|g|\rightarrow |g|\left[ 1-\int_{0}^{1}dtD(t)\left( 1-\cos \phi \right)
\right] .
\]
[Analogous renormalization of $|g|$ owing to pair-breaking is known for
proximity-effect S/N-bilayers in the so-called Cooper limit.\cite{dG64}] As
a consequence, for $D\sim \left| g\right| N(0)$, we expect non-sinusoidal
current-phase dependence in the whole temperature range.

On the contrary, for 
\begin{equation}
\int_{0}^{1}dtD(t)<<|g|N(0),  \label{15}
\end{equation}
equation (\ref{10}) gives a fully self-consistent description of the
Josephson current at $T<T^{*}$, whereas the situation in the temperature
range $T^{*}\leq T\leq T_{c0}$ yet needs clarification.

\section{The effect of intrabarrier exchange interaction}

To investigate the effect of intrabarrier exchange interactions, we again
turn to the simplest case of thin insulating barriers with $d<<\min \{a,\xi
_{0}\}$. In this case, the barrier potential can be modeled by\cite{KF92-2} 
\begin{equation}
\hat{U}_{B}(x)=\left( V+J\sigma _{3}\right) \sum_{n=-\infty }^{+\infty
}\delta (x-na),\text{ }V>0,  \label{16}
\end{equation}
where $V$ and $J$ are the nonexchange and exchange parts, respectively.

In the limit of a low barrier transmission, when $V\gg \left| J\right| \gg
v_0$, we can employ the perturbation procedure described in Sec. III,
obtaining

\[
\hat G_\omega ^{(1)}(x\in S_1,x^{\prime }\in S_0;t)= 
\]

\[
=-\frac 1{4\Omega ^2}\frac{V-J\sigma _3}{V^2}\left\{ \left[ (\Omega +\omega
)^2+\Delta _0^2e^{i\phi }\right] \frac{\sin \left[ \lambda _{+}(a-x)\right]
\sin \left[ \lambda _{+}(a+x^{\prime })\right] }{\sin {}^2\lambda _{+}a}%
\right. 
\]

\[
+\Delta _0^2\left( 1-e^{i\phi }\right) \frac{\sin \left[ \lambda
_{+}(a-x)\right] \sin \left[ \lambda _{-}(a+x^{\prime })\right] +\sin \left[
\lambda _{-}(a-x)\right] \sin \left[ \lambda _{+}(a+x^{\prime })\right] }{%
\sin \lambda _{+}a\sin \lambda _{-}a} 
\]

\begin{equation}  \label{17}
+\left. \left[ (\Omega -\omega )^2+\Delta _0^2e^{i\phi }\right] \frac{\sin
\left[ \lambda _{-}(a-x)\right] \sin \left[ \lambda _{-}(a+x^{\prime
})\right] }{\sin {}^2\lambda _{-}a}\right\} .
\end{equation}
[Compare with Eq. (\ref{7})]. The Green's function in the normal state, $%
\hat G_\omega ^{n(1)}$, can be obtained from (\ref{17}) by taking the limit $%
\Delta _0\rightarrow 0$.

Substituting $\Delta ^{(0)}(x)=\Delta _{0}(T)\sum_{n=-1}^{1}\exp (in\phi
)\delta _{S_{n}}(x)$ and quasiclassical approximation for $\hat{G}_{\omega
}^{n(1)}$, $\hat{G}_{\omega }^{(1)}$ into (\ref{3}) with $d=0$, we get 
\begin{equation}
j=\frac{\pi }{2}ev_{0}N(0)\Delta _{0}^{2}T\int_{0}^{1}dtt\left[
D(t)-2D_{S}(t)\right] \sum_{\omega }\frac{\tanh \frac{2a\sqrt{\omega
^{2}+\Delta _{0}^{2}}}{v_{0}t}}{\omega ^{2}+\Delta _{0}^{2}}\sin \phi ,
\label{18}
\end{equation}
where 
\[
D(t)=\frac{v_{0}^{2}\left( V^{2}+J^{2}\right) }{V^{4}},\text{ }\,\text{ }%
D_{S}(t)=\frac{v_{0}^{2}J^{2}}{V^{4}}
\]
are the total tunneling probability and the exchange part of the tunneling
probability, respectively. In Eq. (\ref{18}), the typical difference $%
D-2D_{S}$ can be regarded\cite{KF86,KF95} as the tunneling probability for a
Cooper pair, with $2D_{S}$ being the probability of pair breaking. In the
limit $D_{S}=0$ ($J=0$), equation (\ref{18}) reduces to (\ref{8}), as it
should.

For $a\gg \xi _{0}$, equation (\ref{18}) becomes\cite{KF95} 
\begin{equation}
j=\frac{\pi ev_{0}N(0)}{2}\int_{0}^{1}dtt\left[ D(t)-2D_{S}(t)\right] \Delta
_{0}(T)\tanh \frac{\Delta _{0}(T)}{2T}\sin \phi .  \label{19}
\end{equation}

In the opposite {\it mesoscopic} limit $p_{0}^{-1}\ll a\ll \xi _{0}$, we
obtain 
\begin{equation}
j=\frac{1}{2\pi }\frac{ev_{0}N(0)\Delta _{0}^{2}(T)}{T_{c0}}\frac{a}{\xi _{0}%
}\frac{\int_{0}^{1}dt\left[ D(t)-2D_{S}(t)\right] }{\left| g\right| N(0)}%
\sin \phi .  \label{20}
\end{equation}
This equation implies that $\left( D-2D_{S}\right) /\left| g\right| N(0)\ll 1
$. As in the case of Eq. (\ref{10}), to establish the temperature range of
validity of Eq. (\ref{20}), we should consider the corresponding linearized
equation

\[
\Delta (x\in S_{0})=\frac{\pi N(0)\left| g\right| }{v_{0}}\left[
\int_{-a+0}^{-0}dx^{\prime }\Delta (x^{\prime })T\sum_{\omega }\int_{0}^{1}%
\frac{dt}{t}\left\{ \exp \left[ -\frac{2|\omega |}{v_{0}t}\left| x-x^{\prime
}\right| \right] \right. \right. 
\]

\[
\left. \left[ 1-D(t)-2D_S(t)\right] \frac{\exp \left[ -\frac{2|\omega |a}{%
v_0t}\right] \cosh \left[ \frac{2|\omega |}{v_0t}\left( x-x^{\prime }\right)
\right] +\cosh \left[ \frac{2|\omega |}{v_0t}\left( x+x^{\prime }+a\right)
\right] }{\sinh \frac{2|\omega |a}{v_0t}}\right\} 
\]

\[
+\int_{+0}^{a-0}dx^{\prime }\Delta (x^{\prime })T\sum_{\omega }\int_{0}^{1}%
\frac{dt}{t}\left[ D(t)-2D_{S}(t)\right] \left\{ \frac{\exp \left[ -\frac{%
2|\omega |}{v_{0}t}\left( x^{\prime }-x-a\right) \right] }{2\cosh \frac{%
2|\omega |a}{v_{0}t}}\right. 
\]

\[
\left. +\frac{\exp \left[ -\frac{2|\omega |a}{v_0t}\right] \cosh \left[ 
\frac{2|\omega |}{v_0t}\left( x-x^{\prime }+a\right) \right] +\cosh \left[ 
\frac{2|\omega |}{v_0t}\left( x+x^{\prime }\right) \right] }{\sinh \frac{%
4|\omega |a}{v_0t}}\right\} 
\]

\[
+\int_{-2a+0}^{-a-0}dx^{\prime }\Delta (x^{\prime })T\sum_{\omega
}\int_{0}^{1}\frac{dt}{t}\left[ D(t)-2D_{S}(t)\right] \left\{ \frac{\exp
\left[ -\frac{2|\omega |}{v_{0}t}\left( x-x^{\prime }-a\right) \right] }{%
2\cosh \frac{2|\omega |a}{v_{0}t}}\right. 
\]

\begin{equation}  \label{21}
\left. \left. +\frac{\exp \left[ -\frac{2|\omega |a}{v_0t}\right] \cosh
\left[ \frac{2|\omega |}{v_0t}\left( x-x^{\prime }-a\right) \right] +\cosh
\left[ \frac{2|\omega |}{v_0t}\left( x+x^{\prime }-2a\right) \right] }{\sinh 
\frac{4|\omega |a}{v_0t}}\right\} \right] .
\end{equation}

In the limit $a>>\xi _{0}$, equation (\ref{21}) goes over into the equation
for a single junction with a tunnel magnetic barrier,\cite{KF86} describing
spatial dependence of $\Delta $ near $T_{c0}$ in the vicinity of the
barrier. For $D_{S}=0$, equation (\ref{21}) reduces to (\ref{11}), as
expected. But by contrast to Eq. (\ref{11}), equation (\ref{21}) has
non-trivial solution at $T<T_{c0}$ even for real $\Delta (x)=\Delta
(T)\sum_{n=-1}^{1}\delta _{S_{n}}(x)$ (i. e., in the absence of the
supercurrent). These solutions determine the critical temperature $T_{c}$: 
\begin{equation}
T_{c}=T_{c0}\exp \left[ -\frac{4}{|g|N(0)}\int_{0}^{1}dtD_{S}(t)\right] .
\label{22}
\end{equation}
Expression (\ref{22}) has a standard form of the critical temperature in the
presence of non-ergodic pair-breaking interactions, typical of small-size
(on the scale $\xi _{0}$) superconducting systems.\cite{dG89} As we see, for 
$D_{S}\sim \left| g\right| N(0)$, the temperature shift due to intrabarrier
exchange interactions in mesoscopic superlattices cannot be disregarded,
whereas in the limit $a\gg \xi _{0}$, with $D_{S}\not{=}0$, one would have%
\cite{KF86} $T_{c}=T_{c0}$.

Substituting now $\Delta (x)=\Delta (T)\sum_{n=-1}^{1}\exp (in\phi
_{c})\delta _{S_{n}}(x)$, we determine the critical phase difference $\left|
\phi _{c}\right| $: 
\begin{equation}
\left| \phi _{c}\right| =\arccos \left[ 1+\frac{|g|N(0)}{\int_{0}^{1}dt%
\left[ D(t)-2D_{S}(t)\right] }\ln \frac{T}{T_{c}}\right] ,  \label{23}
\end{equation}
where $T^{*}\leq T\leq T_{c}$, with $T^{*}$ given by 
\begin{equation}
T^{*}=T_{c}\exp \left[ -\frac{1}{|g|N(0)}\int_{0}^{1}dt\left[
D(t)-2D_{S}(t)\right] \right] .  \label{24}
\end{equation}

At temperatures $T<T^{*}$, the superconductivity of the S-layers cannot be
destroyed completely by the Josephson current, although the suppression of
the gap parameter is not necessarily small. For example, at $T=0$ the gap is
given by 
\begin{equation}
|\Delta (0)|=\Delta _{0}(0)\exp \left[ -\frac{1}{|g|N(0)}\int_{0}^{1}dt%
\left[ D(t)\left( 1-\cos \phi \right) +2D_{S}\left( 1+\cos \phi \right)
\right] \right] .  \label{25}
\end{equation}
Thus, combined pair-breaking effect of the intrabarrier exchange
interactions and the Josephson current can be regarded as an effective
renormalization of the BCS coupling constant: 
\[
|g|\rightarrow |g|\left[ 1-\int_{0}^{1}dt\left[ D(t)\left( 1-\cos \phi
\right) +2D_{S}\left( 1+\cos \phi \right) \right] \right] .
\]

In the limit 
\begin{equation}
\int_{0}^{1}dt\left[ D(t)-2D_{S}(t)\right] \ll |g|N(0),  \label{26}
\end{equation}
the exponential in Eq. (\ref{25}) can be expanded into a power series, which
proves the self-consistency of Eq. (\ref{20}) for $T<T^{*}$.

\section{Summary and conclusions}

Summarizing, we have discussed some of the basic theoretical aspects of
current-carrying states in superconducting mesoscopic superlattices with
tunnel barriers.

In particular, we have derived self-consistent analytical expressions for
the Josephson current in these structures [Eqs. (\ref{10}) and (\ref{20})],
valid under conditions (\ref{15}) and (\ref{26}), respectively, at
temperatures $T<T^{*}$, with $T^{*}$ given by (\ref{13}) and (\ref{24}). We
have explained the peculiarities of Eqs. (\ref{10}) and (\ref{20}), i. e., a
strong depression of $j_c$ due to the factor $a/\xi _0$ and unusual
temperature dependence given by $\Delta _0^2(T)$, in terms of nonlocality
inherent to the theory of superconductivity. We have shown that in the
temperature range $T^{*}\leq T<T_c$ the Josephson current can completely
destroy the superconductivity of the S-layers at a certain critical phase
difference $\left| \phi _c\right| $ [Eqs. (\ref{12}) and (\ref{23})]. Our
equations (\ref{14}), (\ref{25}) reveal an exponential decrease of the gap
parameter at $T=0$ due to the Josephson current, unless the conditions (\ref
{15}), (\ref{26}) are fulfilled. Moreover, equations for the critical
temperature (\ref{22}) and the gap parameter at $T=0$ (\ref{25}) in the case
of magnetic tunnel barriers clearly demonstrate that, by contrast to the
single-junction case, the influence of intrabarrier exchange interactions
cannot be completely disregarded in the mesoscopic superlattices too.

The above results may have important practical implications for
superconducting low-$T_c$ technology. For example, the predicted effect of a
strong reduction of $j_c$ in the mesoscopic regime establishes a
quantitative limit on decreasing the S-layer thickness in vertically-stacked
Josephson-junction arrays intended for superconducting microelectronic
circuitry of high integration.

Finally, we cannot but mention some of unresolved problems. As we have
already pointed out, the situation in the temperature range $T^{*}\leq T<T_c$
needs further clarification, as well as the effect of nonmagnetic impurities
in the S-layers. Another interesting issue is the combination of the above
discussed effects, resulting from spatial nonlocality, with anisotropic ($d$%
-wave) pairing now proposed for high-$T_c$ superconductors. All this
suggests that further theoretical and experimental studies of weakly-coupled
mesoscopic superlattices would be highly desirable.

\bigskip\ 

\begin{center}
{\bf ACKNOWLEDGMENTS}
\end{center}

\smallskip\ The authors gratefully acknowledge stimulating discussions with 
\v S. Be\v na\v cka, U. Gunsenheimer, A. N. Omel'yanchuk, S. Tak\'acs, and
I. V\'avra.

\end{document}